\documentclass[preprint, 12pt]{elsarticle}

\usepackage{graphicx}
\usepackage[T1]{fontenc}
\usepackage{amsmath}
\usepackage{amsfonts}
\usepackage{float}
\usepackage{xcolor}
\usepackage{soul}

\usepackage{amssymb}
\journal{\tt arXiv}

\begin{document}

\begin{frontmatter}



\title{Prediction and Retrodiction in Statistical Mechanics from the Principle of Maximum Caliber}


\author[dfc]{Ignacio Tapia}
\ead{ign.tap@outlook.com}
\author[dfc]{Gonzalo Guti\'errez}
\ead{gonzalogutierrez@uchile.cl}
\author[p2mc]{Sergio Davis}
\ead{sergdavis@gmail.com}
\affiliation[dfc]{organization={Departamento de F\'isica, Facultad de Ciencias, Universidad de Chile},%Department and Organization
            addressline={Las Palmeras 3425, \~Nu\~noa}, 
            city={Santiago},
            postcode={}, 
            state={},
            country={Chile}
            }

\affiliation[p2mc]{organization={Centro de Investigaci\'on en la Intersección entre F\'isica de Plasmas, Materia y Complejidad (P$^2$mc), Comisi\'on Chilena de Energ\'ia Nuclear},%Department and Organization
            addressline={Casilla 188-D}, 
            city={Santiago},
            postcode={}, 
            state={},
            country={Chile}
            }

\begin{abstract}
%me parece que lo que se propone no es una int3erpretacion, sino un marco conceptual completo para obtener modelos (o teorías como le llaman en Beaucheff
A statistical, path-dependent framework to describe time-dependent macroscopic theories using the Principle of Maximum Caliber is presented. By means of this procedure, it is possible to infer predictive non-equilibrium statistical mechanical models from a variational principle, provided that the adequate time-dependent constraints and the state of the system at some specific times are given.
The approach is exemplified by obtaining the description of a time-dependent Brownian particle from kinetic restrictions.
We relate the predictive nature of a model to the structure of the prior distribution that represents the state of knowledge about the system before the dynamical constraints are considered.
Non-predictive models are shown to be possible in the presented framework and as an example, retrodictive dynamics are obtained from the same kinetic constraints.
\end{abstract}


%\item Prediction and retrodiction in stochastic systems from variational principle
%\item Time-dependent stochastic dynamics from the Principle of Maximum Caliber
%\item Time-dependent normal diffusion from kinetic constraints and entropy maximization
%\item Time-dependent macroscopic states as a marginalization of path distribution


\end{frontmatter}


\section{Introduction}

The aim of Statistical Mechanics, as the Encyclopedia Britannica states,
%OJO: o poner aqui cualquier otra definicion de librosreconocidos
is ``to predict
and explain the measurable properties of macroscopic systems on the basis of the microscopic constituents of those systems''. Due to the large number of
particles (``microscopic constituents'') involved, one needs to make use of probability theory and statistics to  \emph{estimate} the macroscopic quantities in the form of expectations.
Equilibrium Statistical Mechanics, in particular, follows this program, and was put in firm basis thanks to the works of Boltzmann, Gibbs and others at the beginning of the last century \cite{Landau}.
%Poner ref. de libros conocidos: propongo Landau de Statistical physics o Huang, o Ma)
Their hypothesis, useful to characterize the macroscopic state of systems in equilibrium with constraints, can be summarized in what is known as the Principle of Maximum Entropy (MaxEnt).
This variational principle proposed by Jaynes \cite{Jaynes1} as a generalization of Boltzmann and Gibbs ideas,
makes it possible to estimate the macroscopic behavior of a system from specific given information.

Non-Equilibrium Statistical Mechanics (NESM), on the other hand,  deals with macroscopic states that depend on time \cite{Eu1998-bg}.
Although there is no a general, accepted theory so far, 
the usual approach for the study of these regimes consist of \emph{predicting} how a probability distribution evolves from a known initial condition.
Several schemes have been put forward to study the time evolution of probability density functions.
%OJO: aqui quieres decir como evoluciona la forma funciona de la probabilidad que depende del tiempo, o como evoluciona la probabilidad en el tiempo? 
% En este ultimo cao seria: "Several schemes have been put forward to study the time evolutiuon fo the probability density functions"
This kind of approach treats time as a label unrelated to the microscopic states, which are considered as random variables.
The Principle of Maximum Caliber (MaxCal), proposed also by Jaynes \cite{Jaynes2},
has been shown to be a generalization of MaxEnt for dynamical random variables
that can be used to estimate the time-dependent behavior of macroscopic quantities from known constraints.
%Aqui de nuevo esta confuso lo que quieres decir, muy larga la frase: favor ponlo en castellano y lo traducimos. Me explico: que es lo que puede ser usado para estimar el time-dependent...? el MaxCal, o las random variables?)
This principle has emerged as a strong candidate for a description of NESM. In fact, in the last ten years the number of practitioners has augmented, and several
recent excellent reviews exist, mainly by the group of Ken Dill and collaborators \cite{Caliber1, Caliber2}. There, it can be seen that MaxCal has been used to describe
different previously reported models in a more transparent way \cite{Caliber3, Caliber5, CaliberApp2, Davis2018}, as well as to other, new problems, even beyond physics \cite{Caliber4, CaliberApp1, CaliberApp3}.

In this paper, we show explicitly how the Principle of Maximum Caliber can be used to describe time-dependent, macroscopic systems.
We focus on providing a description for predictive NESM theories, but we also show that the path-dependent formalism allows for the characterization of more general, non-predictive macroscopic theories.
We do this by studying the case of retrodiction, a type of estimation where the macroscopic state of the system is inferred from information about its future.
Although less common than prediction, retrodiction has been studied both in statistical physics \cite{RetroStat1, RetroStat2, RetroStat3} and in quantum mechanics \cite{RetroQM1, RetroQM2, RetroQM3}.

%COMMENT: esto que sigue ponerlo como nota en la referencia: By the term "Estimation" we refer to results obtained as expectation value from a probabilistic theoretical model; the word prediction refer to estimations concerning the future. Retrodiction refer to estimations concerning the past

\section{Time-dependent probabilities}

Among the macroscopic theories to study time-dependent probability distributions, perhaps the best known are the Kinetic Theory of gases and the formalism of Stochastic Processes.
These theories can be used to describe different macroscopic regimes for time-dependent systems.
A system is completely described when the time-dependent probability distribution $\rho(x \vert t)$ is specified. Once the distribution is found for every time $t$, it can be used to estimate macroscopic properties of the system in the form of expectations

\begin{equation}
\left \langle f(x) \right \rangle_t = \int f(x) \rho(x \vert t) \, dx \, \text{.}
\end{equation}

A specific theory must provide an expression for the macroscopic properties $f(x)$ as a function of the microscopic degrees of freedom, but it also must propose a model for the temporal behavior of the distribution $\rho(x \vert t)$. In general, theories in NESM are predictive. This means that they can be used to estimate the time-dependent distribution $\rho(x \vert t)$ from an initial condition $\rho(x \vert t = 0) = \rho_0(x)$. The initial condition is arbitrary. It cannot be obtained from the model and only when it is specified (from experimental observation or physical suppositions) the theory leads to a complete description of the system. Therefore the model and the initial condition can be thought of as different kinds of knowledge. It is important to distinguish between these two aspects that are common to every NESM theory. This is because the model is a consequence of the theory and is related to general physical laws, while the initial condition is intrinsic to each system and has to be known or assumed from observation.

For evolution equations that are linear on the probability density function, such as the ones derived from a continuity equation, the dynamics can be condensed into a master equation
\begin{equation} \label{master}
\rho(x \vert t) = \int P(x ; x^{\prime} \vert t) \rho(x^{\prime} \vert 0) \, d x^{\prime} \, \text{,}
\end{equation}
that is, the probability distribution $\rho$ that represents the state of the system at time $t$ can be written as a linear functional of the initial conditions.
The kernel $P(x ; x^{\prime} \vert \Delta t)$ is known as the transition probability, and is the solution for the system with localized initial conditions $P(x ; x^{\prime} \vert 0) = \delta(x - x^{\prime})$. The transition probability, once found, can be used to construct the general solution for the dynamics.
This means the transition probability characterizes the model.

For some theories, the transition probability can be written as a path integral \cite{Kleinert2004-zx} that can be related to the differential equation of the system by the Feynman-Kac formula \cite{Del_Moral2004-jf}.
Path integrals or functional integrals provide a mathematical tool to characterize time-dependent macroscopic states.
This formulation shows a route for an ensemble perspective to describe NESM where the temporal dependence of the system is described in the microscopic degrees of freedom instead of described as a macroscopic parameter. 

\section{Path-dependent probabilities}

The macroscopic behavior of a system follows from the description of its microscopic constituents. In the case of mechanical systems, this description is usually written in a differential language, providing an evolution equation for the microscopic degrees of freedom that acts as a constraint to the dynamics of the macroscopic properties.
The time-dependent statistical description shown in the previous section does not provide an explicit representation of these microscopic dynamics, because the microstates do not depend explicitly on time. The lack of a dynamical representation of the microscopic random variables obscures the identification of the constraints that induce a given macroscopic behavior. Also, the representation of some macroscopic properties is different for different time-dependent theories, as for example the flux (or current density), for which a proper microscopic representation would involve derivatives of the microstates.

In order to construct a dynamical model from constraints, we will consider a statistical description,
but instead of having a time-dependent probability distribution characterizing the system,
we will consider time as a parameter of the microscopic states.
We will consider the random variable of this framework to be a function of time $\tilde{x}$ that for each time $t$ takes a definite value $\tilde{x}(t) = x$.
A time-dependent random variable of this kind will be called a path.

The dynamics of the system will be characterized by assigning a weight $\varrho[\tilde{x}]$, in fact a path probability density, to each possible path $\tilde{x}$.
Time-dependent expectations of observables $f[\tilde{x}]$ can be computed as path integrals in the form
\begin{equation}
\left \langle f(t) \right \rangle = \int \varrho[\tilde{x}] f[\tilde{x}]\, \mathcal{D} \tilde{x} \, \text{.}
\end{equation}
The observable inside the expectation can depend on the path evaluated at some specific time $f[\tilde{x}] = g(\tilde{x}(t))$,
but in general it is a functional of the path.

The advantages of the path integral formalism in comparison with the time-dependent statistics perspective relies on the direct relation between the microscopic and macroscopic variables of the system.
In quantum mechanics, this perspective avoids the operator representation of observables, as they can be expressed directly in the path representation as a function of phase space variables.
In NESM, the same can be done to compute expectations of quantities that do not appear in the usual time-dependent description.
Moreover, the characterization of the dynamics using a single non-dynamical representation $\varrho$ enables the construction of the different models from a variational principle.

\section{Marginalization and time-dependent macroscopic description}

The path-dependent probability density defined in the previous section assigns a weight to every possible path $\tilde{x}$.
Using this distribution,
we can construct a time-dependent probability distribution
that corresponds to the representation of the macroscopic state used in the literature.
This can be done by the process of marginalization and can be written using the expectation of a delta function
\begin{equation} \label{eq:slice}
\rho(x \vert t) = \int \varrho[\tilde{x}] \delta(\tilde{x}(t) - x) \, \mathcal{D} \tilde{x} \, \text{.}
\end{equation}

Equation \eqref{eq:slice} can be read as follows: the probability distribution that describes the behavior of measuring $x$ at time $t$ is equal to the sum of the probability of every path $\tilde{x}$ that is compatible with that measurement, i.e., every path that satisfies $\tilde{x}(t) = x$.

The normalization of the time-dependent probability density $\rho$ is inherited from the normalization of the path-dependent probability density $\varrho$. Also, from the definition it can be proved~\cite{Davis2018} that it satisfies the continuity equation
\begin{equation}
\frac{\partial \rho}{\partial t} = - \frac{\partial J}{\partial x} \, \text{,}
\end{equation}
where the flux $J$ can be expressed also as an expectation over paths
\begin{equation}
J(x \vert t) = \int \varrho[\tilde{x}] \dot{\tilde{x}}(t) \delta(\tilde{x}(t) - x) \, \mathcal{D} \tilde{x} \, \text{.}
\end{equation}
In the time-dependent description, to provide an explicit representation for the flux, it is necessary to specify a theory to model the temporal behavior of the system. The expression found here is universal to every Non-Equilibrium theory.

In the path-dependent description, both the time-dependent density and the flux can be expressed as functions of the path distribution $\varrho$.
This implies that the path distribution contains not only the information about the physical laws that govern the behavior of the system, but it also contains the information about the initial conditions of the system.

\section{Macroscopic Constraints and the Principle of Maximum Caliber}

We will consider the macroscopic constraints to be time-dependent expectations of the form
\begin{equation} \label{constrains}
\left \langle f_n[\tilde{x}](t) \right \rangle = \bar{f}_n(t) \, \text{,}
\end{equation}
%Que es "n". Explicarlo.
where the index $n$ distinguishes between different constraints. We can find several path-dependent distributions $\varrho[\tilde{x}]$ that match these constraints. The Principle of Maximum Caliber states that, from all the possible choices $\varrho$ that are compatible with the restrictions, the most unbiased is the one that maximizes the Caliber $C$ given by
\begin{equation} \label{eq:caliber}
C[\varrho] = - \int \varrho[\tilde{x}] \ln{\frac{\varrho[\tilde{x}]}{\pi[\tilde{x}]}} \, \mathcal{D} \tilde{x} \, \text{.}
\end{equation}

The Caliber is a generalization of the Entropy for situations where the random variable is a path.
The definition \eqref{eq:caliber} contemplates a prior distribution $\pi[\tilde{x}]$ that encapsulates our knowledge about the system without considering constraints.
The maximization of the Caliber updates our information from the prior distribution $\pi[\tilde{x}]$ to the posterior distribution $\varrho[\tilde{x}]$ in order to include the constraints.

We can include the normalization condition and the macroscopic constraints \eqref{constrains} with the help of Lagrange multipliers $\lambda$ and $\beta_n(t)$. The Lagrange multipliers $\beta_n(t)$ associated with time-dependent restrictions also depend on time, and must be included as an integral. The auxiliary function to maximize corresponds to 
\begin{equation}
\bar{C}[\varrho] = C[\varrho] - \lambda \left \langle 1 \right \rangle - \sum_n \int \beta_n(t) \left \langle f_n[\tilde{x}](t) \right \rangle \, dt \, \text{.}
\end{equation}
The solution to the Euler-Lagrange equation associated with this functional is given by
\begin{equation}
\varrho[\tilde{x}] = \frac{\pi[\tilde{x}]}{Z} \exp{\left( - \sum_n \int \beta_n(t) f_n[\tilde{x}](t) \, dt \right)} \, \text{,}
\end{equation}
with the partition function $Z$ guaranteeing normalization expressed as
\begin{equation}
Z = \int  \pi[\tilde{x}] \exp{\left( - \sum_n \int \beta_n(t) f_n[\tilde{x}](t) \, dt \right)} \, \mathcal{D} \tilde{x} \, \text{,}
\end{equation}
and the Lagrange multipliers $\beta_n$ related formally to the constraints as
\begin{equation}
\bar{f}_n(t) = - \frac{\delta }{\delta \beta_n(t)} \ln{Z} \, \text{.}
\end{equation}

With the help of the Principle of Maximum Caliber, the choice of constraints determines an unique manner in how the prior distribution is updated. This is analogous to how a specific time-dependent theory updates an initial condition into a time-dependent distribution that can be used to predict macroscopic properties. In this context, the Maximum Caliber Principle directly relates the macroscopic information about a physical regime to the dynamical model that objectively best describes its behavior.

We will think of the prior distribution as a different kind of knowledge about the system than the constraints: the prior distribution will be used to include information about a specific system; the constraints will account for the physical laws that produces a specific time-dependent model. For this reason, in the context of NESM, we will assume a prior distribution equal to the initial conditions of the time-dependent distribution $\pi[\tilde{x}] = \rho_0(\tilde{x}(0))$.

\section{Prior Distribution and Path Integration}

The specific choice of a procedure to solve the path integrals involved in the theory represent another constraint to the system. This choice will provide the system with a specific microscopic description of the structure of paths. Here we will treat path integration as it was first introduced into Quantum Mechanics by Feynman \cite{Feynman}. That is, we will think of paths $\tilde{x}$ as continuous but non-differentiable functions in order to have independence between degrees of freedom associated with different times. If we take the paths to be defined in the interval $[0, T]$, this independence can be written formally as $\mathcal{D} \tilde{x} = \prod_{t=0}^{T} d \tilde{x}(t)$.

%We will use a discretization scheme as the one shown in \cite{Feynman} and \cite{Kleinert2004-zx}. Time is considered a discrete label $t_n$ and the values of the path at those times $x_n = \tilde{x}(t_n)$ can be thought as a continuous random variable. Every functional of the path has to be replaced by a function of the new discrete random variables, and path integrals are interpreted as vector integrals. The continuum limit is retrieved by taking the limit were the discretization interval $t_{n+1} - t_n$ goes to zero. 

Usually path integrals are computed taking the end values of the paths fixed. In the framework proposed here, paths are as general as possible to account for a statistical description. We will not consider fixed end values in the definition of the path integration, but the fixed points will appear in the computation of expectations, due to the nature of the estimations performed.

The predictive nature ensured by the structure of the prior distribution can lead to a description where every expectation has a fixed initial value. Given that the prior distribution $\pi[\tilde{x}] = \rho_0(\tilde{x}(0))$ depends only on the initial values of paths, we can introduce a delta function $\delta(x^{\prime} - \tilde{x}(0))$ to write it as
\begin{equation}
\pi[\tilde{x}] = \int \rho_0(x^{\prime}) \delta(x^{\prime} - \tilde{x}(0)) \, dx^{\prime} \, \text{.}
\end{equation}
Noting that $x^{\prime}$ is an introduced parameter, independent to the paths, it does not participate in any expectation. This means that every expectation can be obtained from the path distribution with a localized prior distribution $\pi[\tilde{x}] = \delta(x^{\prime} - \tilde{x}(0))$. To include arbitrary initial conditions, the expectations obtained have to be convolved with this initial condition according to the marginalization rule~\cite{Sivia2006},
\begin{equation}
\left \langle f[\tilde{x}] \right \rangle = \int \left \langle f[\tilde{x}] \right \rangle_{\tilde{x}(0) = x^{\prime}} \rho_0(x^{\prime}) \, dx^{\prime} \, \text{,}
\end{equation}
where $\left \langle \ldots \right \rangle_{\tilde{x}(0)=x^{\prime}}$ denotes an expectation taken over paths with fixed initial values $\tilde{x}(0) = x^{\prime}$. In particular, the time-dependent distribution can be written as
\begin{equation} \label{cal_master}
\rho(x \vert t) = \left \langle \delta(\tilde{x}(t) - x) \right \rangle = \int \left \langle \delta(\tilde{x}(t) - x) \right \rangle_{\tilde{x}(0) = x^{\prime}} \rho_0(x^{\prime}) \, dx^{\prime} \, \text{.}
\end{equation}
Equation \eqref{cal_master} is the same as equation \eqref{master}, implying that the transition probability of the model can be written as a path integral with two fixed points, corresponding to
\begin{eqnarray}
P(x;x^{\prime} \vert t) & = & \left \langle \delta(\tilde{x}(t) - x) \right \rangle_{\tilde{x}(0) = x^{\prime}} \\
& = & \frac{1}{Z} \int_{\tilde{x}(0) = x^{\prime}}^{\tilde{x}(t) = x} \exp{\left( -\sum_n \int_{0}^T \beta_n(t) f_n[\tilde{x}](t)  \right)} \, \mathcal{D} \tilde{x} \, \text{,}
\end{eqnarray}
where the fixed values were explicitly included in the path integral.

The predictive nature of the models constructed relies on the structure of the prior distribution. We say it is predictive, because it depends only on the initial values of paths, thus the estimations performed concern the future of the system. Nevertheless, we will show that other structures can be chosen for the prior distribution to obtain non-predictive theories.

We have shown here how to include the time-dependent initial conditions in a path-dependent statistical description of NESM. The information that accounts for the temporal behavior of the system, necessary to complete a description and to perform predictions, must be included in the specific choice of constraints.

\section{Example: Free Particle in a heat bath}

We will consider a one-dimensional free particle of mass $m$ in the presence of a heat bath of temperature $\tau(t)$. The heat bath will be introduced in the macroscopic description of the particle as a constraint on the kinetic energy
\begin{equation}
\frac{m}{2} \left \langle \dot{\tilde{x}}^2(t)  \right \rangle = f(t) \, \text{.}
\end{equation}
If we maximize the Caliber considering this expectation as a restriction, we obtain the path probability distribution given by
\begin{equation} \label{eq:pathdis}
\varrho[\tilde{x}] = \frac{\pi[\tilde{x}]}{Z[\beta]} \exp{\left( - \frac{m}{2} \int \beta(t) \dot{\tilde{x}}^2(t) \, dt  \right)} \, \text{.}
\end{equation}
The partition function can be written as
\begin{equation}
Z[\beta] = \int \exp{\left( - \frac{m}{2} \int \beta(t) \dot{\tilde{x}}^2(t) \, dt \right)} \pi[\tilde{x}] \, \mathcal{D} \tilde{x} \, \text{,}
\end{equation}
and the expectation value of the kinetic energy relates formally to the partition function by
\begin{equation} \label{eq:const}
\frac{m}{2} \left \langle \dot{\tilde{x}}^2(t) \right \rangle = - \frac{\delta}{\delta \beta(t)} \ln{Z[\beta]} \, \text{.}
\end{equation}

The most probable paths satisfy the differential equation
\begin{equation} \label{maximum}
\frac{d}{d t} \left( \beta(t) \dot{\tilde{x}}(t) \right) = 0 \, \text{,}
\end{equation}
which has an infinite amount of solutions without providing boundary conditions.

The Lagrange multiplier $\beta(t)$ will not be adjusted using the expectation restricted,
as we will see that this expectation diverges.
However, we will see that this parameter can be related to the temperature of the heat bath $\tau$ through the dynamics of the time-dependent marginalized distribution.

If we take a localized predictive prior $\pi[\tilde{x}] = \delta(\tilde{x}(0) - x^{\prime})$, then the posterior distribution $\varrho$ obtained in \eqref{eq:pathdis} corresponds to a Gaussian distribution on the velocity of the particle. The mean velocity is zero, and the values of the velocity at different times are uncorrelated, satisfying the relation

\begin{equation} \label{eq:corr}
\left \langle \dot{\tilde{x}}(t) \dot{\tilde{x}}(t^{\prime}) \right \rangle_{\tilde{x}(0) = x^{\prime}} = \frac{\delta(t - t^{\prime})}{m \beta(t)} \, \text{.}
\end{equation}

With localized initial conditions $\tilde{x}(0) = x^{\prime}$, the most probable path is the constant solution $\tilde{x}(t) = x^{\prime}$. It satisfies the differential equation \eqref{maximum} with the least possible value for the kinetic energy.

The path integrals are solved by taking a time-slice approach, as shown in the majority of textbooks (see \cite{Kleinert2004-zx}, chapter 2). By taking partitions $\left \{ t_n = n T  / N \right\}_{0 <= n <= N}$ of the interval $[0, T]$ we can express the path integral as the limit of a $N$-dimensional vector integral. The result obtained corresponds to

\begin{equation} \label{eq:part}
Z[\beta] = \lim_{N \to \infty} \sqrt{ \prod_{n = 1}^{N} \frac{\pi T}{m N \beta(t_n)} } \, \text{.}
\end{equation}

In the continuum limit $N \rightarrow \infty$, the partition function goes to zero.
This implies that the path distribution diverges for every path.
As neither the partition function nor the path distribution are measurable quantities,
we will not ask for these aspects of the theory to converge.

The expectation value of the kinetic energy can be computed directly from the discrete expression \eqref{eq:part}, using the relation \eqref{eq:const} to obtain
\begin{equation}
\frac{m}{2} \left \langle \dot{\tilde{x}}^2(t) \right \rangle_{\tilde{x}(0)=x^{\prime}} = \lim_{N \to \infty} \frac{N}{T} \frac{1}{2 \beta(t)} \, \text{.}
\end{equation}
This quantity again is not convergent in the continuum limit.
Nevertheless, this divergence corresponds to a representation of the delta function $\delta(0) = \lim_{N \to \infty} N / T$, and is consistent with the correlations of velocities from equation \eqref{eq:corr}.

Although the Lagrange multiplier $\beta(t)$ cannot be estimated from this relation,
as there is a divergence,
it can be used to justify that a constant temperature in the heat bath $\tau(t) = \tau_0$ corresponds to a constant value for $\beta(t) = \beta_0$.
We will still consider its temporal dependence to provide a more general approach.

The time-dependent marginalization, for localized initial conditions, corresponds to the transition probability for the predictive system.
It was computed using the same discrete scheme as with the partition function and was found to be finite in the continuum limit, corresponding to
\begin{equation} \label{pred_trans}
P(x ; x^{\prime} \vert t) = \frac{1}{\sqrt{4 \pi \int_0^t dt^{\prime} / m \beta(t^{\prime})}} \exp{\left( - \frac{\left( x - x^{\prime} \right)^2}{2 \int_{0}^{t} dt^{\prime}/m \beta(t^{\prime})} \right)} \, \text{.}
\end{equation}

This is a Gaussian distribution, with constant mean (the same as the most probable path) and an increasing variance, given by
\begin{equation}
\sigma^2(t) = \frac{1}{m} \int_0^t \frac{dt^{\prime}}{\beta(t^{\prime})} \, \text{.}
\end{equation}

The transition probability \eqref{pred_trans} is the kernel for a time-dependent diffusive system.
The system can be interpreted as a time-dependent Brownian motion model, related to the Gaussian processes used in statistical learning.
The parameter $\beta(t)$ can be adjusted from the evolution of the width of an arbitrary initial condition.
The width always grows with time for positive $\beta$
and it depends only on information about the past, as the sum does only involve values of $\beta$ evaluated at times smaller than $t$.

The Lagrange multiplier $\beta$ is related to the diffusion coefficient $D$ of the system
by the expression $\beta = 1/ m D$. In this way,  the Lagrange multiplier is linked to the temperature through Einstein's relation.

The general solution $\rho(x \vert t)$ with arbitrary initial conditions $\rho(x \vert t= 0) = \rho_0(x)$ can be written as a convolution given by equation \eqref{cal_master} and satisfies the time-dependent diffusion equation
\begin{equation}
\frac{\partial \rho}{\partial t} = \frac{1}{m \beta(t)} \frac{\partial^2 \rho}{\partial x^2} \, \text{.}
\end{equation}

\section{Retrodiction in Macroscopic Systems}

As we mentioned, the structure of the prior leads to different types of estimations. Prediction is the case when the estimations are made from information about the past. Explicitly, in this case, the prior distribution depends only on the paths evaluated at an initial time $\tilde{x}(0)$. If we choose the prior distribution to depend only on the values of paths at final times $\tilde{x}(T)$ instead, we obtain a backwards process identical to the predictive process, but with reversed time. This backward process does not describe a physical situation, regardless it is consistent with the prior information assumed.

The procedure to make inference about the past with information about the present or the future is called retrodiction and has been studied in the last decades. It is not obtained by simply taking a final condition instead of an initial condition for the time-dependent distribution $\rho(x \vert t)$. This is because the causal assumption that the system evolves from a definite known initial condition is not only a consideration about our knowledge, but also a statement about the behavior of physical systems: macroscopic time-dependent systems evolve from the past to the future, regardless of our knowledge. If we include information about the final state of the system, we have to also include information about an initial state. 

Choosing a prior distribution that depends on the initial and final values of paths $\pi[\tilde{x}] = \pi_0(\tilde{x}(0), \tilde{x}(T))$, we obtain retrodictive models. Taking the expectations over paths with fixed boundary values $\tilde{x}(0) = x^{\prime}$ and $\tilde{x}(T) = x^{\prime \prime}$, we can write every expectation as a convolution

\begin{equation}
\left \langle f[\tilde{x}] \right \rangle = \int \int \left \langle f[\tilde{x}] \right \rangle_{\tilde{x}(0) = x^{\prime}}^{\tilde{x}(T) = x^{\prime \prime}} \pi_0(x^{\prime}, x^{\prime \prime} ) \, dx^{\prime} \, dx^{\prime \prime} \, \text{,}
\end{equation}
where the expectation $\left \langle \ldots \right \rangle_{\tilde{x}(0) = x^{\prime}}^{\tilde{x}(T) = x^{\prime \prime}}$ has to be taken considering a localized prior $\pi[\tilde{x}] = \delta(\tilde{x}(0) - x^{\prime}) \delta(\tilde{x}(T) - x^{\prime \prime})$. The time-dependent marginalization $\left \langle \delta(\tilde{x}(t) - x) \right \rangle_{\tilde{x}(0) = x^{\prime}}^{\tilde{x}(T) = x^{\prime \prime}}$ acts again as a transition probability, but this time the convolution has to be taken over the joint probability $\pi_0(x^{\prime}, x^{\prime \prime})$. In the case where our knowledge about initial and final state are independent, this joint distribution takes the form $\pi_0(x^{\prime}, x^{\prime \prime}) = \rho(x^{\prime} \vert 0) \rho(x^{\prime \prime} \vert T)$.

\subsection{Retrodictive estimations for the free particle example}

We showed that if we take as a constraint the expectation value of the kinetic energy, we obtain the path-dependent distribution given by equation \eqref{eq:pathdis}. Taking a retrodictive localized prior distribution $\pi[\tilde{x}] = \delta(\tilde{x}(0) - x^{\prime}) \delta(\tilde{x}(T) - x^{\prime \prime})$, we computed again the same quantities that we computed previously for a predictive system.

The partition function can be expressed as a limit given by
\begin{equation}
Z[\beta] = \lim_{N \to \infty} \sqrt{ \prod_{n = 1}^{N-1} \frac{\pi T}{m N \beta(t_n)} } \exp{\left( - \frac{\left( x^{\prime \prime} - x^{\prime} \right)^2}{2 \sum_{n =1}^{N} \Delta t / m \beta(t_n)} \right)}\, \text{.}
\end{equation}
This quantity diverges in a similar manner to the predictive partition function. The expectation value of the kinetic energy, computed from relation \eqref{eq:const} is given by
\begin{equation}
\frac{m}{2} \left \langle \dot{\tilde{x}}^2(t) \right \rangle_{\tilde{x}(0) = x^{\prime}}^{\tilde{x}(T) = x^{\prime \prime}} = \frac{\delta(0)}{2 \beta(t)} + \frac{m}{2} \left( \int_0^{T} \frac{dt^{\prime}}{\beta(t^{\prime})} \right)^{-2} \frac{\left( x^{\prime \prime} - x^{\prime} \right)^2}{\beta^2(t)} \, \text{,}
\end{equation}
which has the divergent term that arises from the uncorrelation between velocities,
but also has a finite part corresponding to the kinetic energy of the solution $\bar{x}(t)$ of equation \eqref{maximum}
with the boundaries given by the localization condition $\bar{x}(0) = x^{\prime}$ and $\bar{x}(T) = x^{\prime \prime}$. This solution can be written as
\begin{equation}
\bar{x}(t) = \frac{x^{\prime} \int_{t}^{T} dt^{\prime}/\beta(t^{\prime}) + x^{\prime \prime} \int_{0}^{t} dt^{\prime} / \beta(t^{\prime})  }{\int_{0}^T dt^{\prime} / \beta(t^{\prime})} \, \text{,}
\end{equation}
and corresponds to the most probable path of this model.

The transition probability  is found to be finite and corresponds to a Gaussian distribution with mean equal to $\bar{x}(t)$ and variance consistent with the localized retrodictive assumptions, as it is zero at boundary times $t = 0$ and $t = T$. The variance or mean square displacement of this distribution is given by
\begin{equation}
\sigma^2(t) = \frac{m}{4} \frac{\left( \int_0^t dt^{\prime} / \beta(t^{\prime}) \right) \left( \int_t^T dt^{\prime}/\beta(t^{\prime}) \right)}{\int_0^T dt^{\prime}/\beta(t^{\prime})} \, \text{.}
\end{equation}
We see that the integrals in the numerator go to zero for initial and final times, guaranteeing the certainty in the initial and final values of paths.

A general solution with initial condition $\rho(x \vert 0) = \rho_0(x)$ and final condition $\rho(x \vert T) = \rho_T(x)$, can be written as a convolution
\begin{equation}
\rho(x \vert t) = \int \int \left \langle \delta(\tilde{x}(t) - x) \right \rangle_{\tilde{x}(0) = x^{\prime}}^{\tilde{x}(T) = x^{\prime \prime}} \rho_0(x^{\prime}) \rho_T(x^{\prime \prime}) \, dx^{\prime} \, dx^{\prime \prime} \, \text{.}
\end{equation}

The retrodictive time-dependent dynamics presented in this section do not correspond to a physical system. The time-dependent distribution obtained interpolates the evolution of a distribution into another one by assuming the dynamic is the one of a Brownian motion. Nevertheless, the dynamics shown here could be useful to account for fluctuations and uncertainty in numerical data analysis. For example, in systems of few particles, where the solutions often deviate from the predictive theory.

\section{Conclusions}

The Principle of Maximum Caliber was presented as a comprehensive, general procedure to
deal with non equilibrium statistical mechanics systems. As a working example, we
applied it to study in detail the case of a diffusive system.
The presence of a heat bath is included as a constraint to the dynamics of the model,
in the form of the expectation value of the kinetic energy of the system.
The structure of the prior distribution, inherited from our knowledge about the system, is responsible for the predictive nature of the estimations of a given model.

The path-dependent formalism is a clearer representation of macroscopic systems than the time-dependent statistics framework. Specifically in the description of the microscopic degrees of freedom involved and their relation with macroscopic properties.
It can be used to write non-equilibrium, statistical theories in a novel manner that provides insight on the relation between macroscopic and microscopic dynamics.
The Principle of Maximum Caliber can be interpreted as a link between the dynamic of a macroscopic system and the specific restrictions that originate that dynamic.

The predictive model presented here is consistent with the results obtained from the usual time-dependent statistical mechanics; dynamics are irreversible and the information about the system can be condensed in a transition probability that can be used to obtain solutions with arbitrary initial conditions.
Kinetic constraints were associated with the thermal dynamics of Brownian Motion.
For arbitrary choice of the parameter $\beta(t) > 0$, the mean square displacement of the macroscopic distribution
is always increasing with time, as it is a sum of positive quantities.
In the case where $\beta$ does not depend on time, we retrieve the normal diffusion process.

Interestingly, the retrodictive model found is also consistent with previously reported models, as for example the Aharonov-Bergmann-Lebowitz rule \cite{RetroQM1} for Quantum Mechanics.

All in all, the Principle of Maximum Caliber, together with the MaxEnt principle, 
represent a new, different theoretical effort to build a 
consistent theory of equilibrium and non equilibrium
statistical mechanics from very few statements about known information, and deserve to be investigated further.

%%%%%%%%%%%%%%%%%%%%%%%%%%%%%%%%%%%%%%%%%%%
%             ACKNOWLEDGMENTS             %
%%%%%%%%%%%%%%%%%%%%%%%%%%%%%%%%%%%%%%%%%%%
\section{Acknowledgments}
The authors thank financial support by the National Agency for Research and Development (Chile)
through 
ANID-PFCHA/Doctorado Nacional/2019-21192159 (I.T.) and grants FONDECYT 1171127 (G.G.) and 
FONDECYT 1220651 (S.D.).




\bibliographystyle{elsarticle-num} 
\bibliography{caliber.bib}

\emailauthor{ign.tap@outlook.com}{Ignacio Tapia}
\emailauthor{gonzalogutierrez@uchile.cl}{Gonzalo Guti\'errez}
\emailauthor{sergdavis@gmail.com}{Sergio Davis}
\Newlabel{dfc}{a}
\Newlabel{p2mc}{b}

\begin{thebibliography}{10}
\expandafter\ifx\csname url\endcsname\relax
  \def\url#1{\texttt{#1}}\fi
\expandafter\ifx\csname urlprefix\endcsname\relax\def\urlprefix{URL }\fi
\expandafter\ifx\csname href\endcsname\relax
  \def\href#1#2{#2} \def\path#1{#1}\fi

\bibitem{Landau}
L.~D. Landau, E.~M. Lifshitz, Statistical Physics, 3rd Edition,
  Butterworth-Heinemann, Oxford, England, 1996.

\bibitem{Jaynes1}
E.~T. Jaynes,
  \href{https://link.aps.org/doi/10.1103/PhysRev.106.620}{Information theory
  and statistical mechanics}, Phys. Rev. 106 (1957) 620--630.
\newblock \href {https://doi.org/10.1103/PhysRev.106.620}
  {\path{doi:10.1103/PhysRev.106.620}}.
\newline\urlprefix\url{https://link.aps.org/doi/10.1103/PhysRev.106.620}

\bibitem{Eu1998-bg}
B.~C. Eu, Nonequilibrium statistical mechanics, 1998th Edition, Fundamental
  Theories of Physics, Springer, Dordrecht, Netherlands, 1998.

\bibitem{Jaynes2}
E.~T. Jaynes, The minimum entropy production principle, Annual Review of
  Physical Chemistry 31~(1) (1980) 579--601.
\newblock \href {https://doi.org/10.1146/annurev.pc.31.100180.003051}
  {\path{doi:10.1146/annurev.pc.31.100180.003051}}.

\bibitem{Caliber1}
K.~Ghosh, P.~D. Dixit, L.~Agozzino, K.~A. Dill, The maximum caliber variational
  principle for nonequilibria, Annual Review of Physical Chemistry 71~(1)
  (2020) 213--238, pMID: 32075515.
\newblock \href {https://doi.org/10.1146/annurev-physchem-071119-040206}
  {\path{doi:10.1146/annurev-physchem-071119-040206}}.

\bibitem{Caliber2}
S.~Press\'e, K.~Ghosh, J.~Lee, K.~A. Dill,
  \href{https://link.aps.org/doi/10.1103/RevModPhys.85.1115}{Principles of
  maximum entropy and maximum caliber in statistical physics}, Rev. Mod. Phys.
  85 (2013) 1115--1141.
\newblock \href {https://doi.org/10.1103/RevModPhys.85.1115}
  {\path{doi:10.1103/RevModPhys.85.1115}}.
\newline\urlprefix\url{https://link.aps.org/doi/10.1103/RevModPhys.85.1115}

\bibitem{Caliber3}
D.~Gonz{\'{a}}lez, S.~Davis, G.~Guti{\'{e}}rrez,
  \href{https://doi.org/10.1007/s10701-014-9819-8}{Newtonian dynamics from the
  principle of maximum caliber}, Foundations of Physics 44~(9) (2014) 923--931.
\newblock \href {https://doi.org/10.1007/s10701-014-9819-8}
  {\path{doi:10.1007/s10701-014-9819-8}}.
\newline\urlprefix\url{https://doi.org/10.1007/s10701-014-9819-8}

\bibitem{Caliber5}
I.~J. General, \href{https://doi.org/10.1103/physreve.98.012110}{Principle of
  maximum caliber and quantum physics}, Physical Review E 98~(1) (Jul. 2018).
\newblock \href {https://doi.org/10.1103/physreve.98.012110}
  {\path{doi:10.1103/physreve.98.012110}}.
\newline\urlprefix\url{https://doi.org/10.1103/physreve.98.012110}

\bibitem{CaliberApp2}
C.~Cafaro, S.~A. Ali,
  \href{https://link.aps.org/doi/10.1103/PhysRevE.94.052145}{Maximum caliber
  inference and the stochastic ising model}, Phys. Rev. E 94 (2016) 052145.
\newblock \href {https://doi.org/10.1103/PhysRevE.94.052145}
  {\path{doi:10.1103/PhysRevE.94.052145}}.
\newline\urlprefix\url{https://link.aps.org/doi/10.1103/PhysRevE.94.052145}

\bibitem{Davis2018}
S.~Davis, D.~González, G.~Gutiérrez, Probabilistic inference for dynamical
  systems, Entropy 20 (2018) 696.

\bibitem{Caliber4}
M.~Otten, G.~Stock, Maximum caliber inference of nonequilibrium processes, The
  Journal of Chemical Physics 133~(3) (2010) 034119.
\newblock \href {https://doi.org/10.1063/1.3455333}
  {\path{doi:10.1063/1.3455333}}.

\bibitem{CaliberApp1}
H.~Wan, G.~Zhou, V.~A. Voelz, \href{https://doi.org/10.1021/acs.jctc.6b00938}{A
  maximum-caliber approach to predicting perturbed folding kinetics due to
  mutations}, Journal of Chemical Theory and Computation 12~(12) (2016)
  5768--5776.
\newblock \href {https://doi.org/10.1021/acs.jctc.6b00938}
  {\path{doi:10.1021/acs.jctc.6b00938}}.
\newline\urlprefix\url{https://doi.org/10.1021/acs.jctc.6b00938}

\bibitem{CaliberApp3}
E.~Johnson, \href{https://onlinelibrary.wiley.com/doi/abs/10.1002/prot.26299}{A
  maximum caliber analysis of the foldon hypothesis}, Proteins: Structure,
  Function, and Bioinformatics 90~(5) (2022) 1170--1178.
\newblock \href
  {http://arxiv.org/abs/https://onlinelibrary.wiley.com/doi/pdf/10.1002/prot.26299}
  {\path{arXiv:https://onlinelibrary.wiley.com/doi/pdf/10.1002/prot.26299}},
  \href {https://doi.org/https://doi.org/10.1002/prot.26299}
  {\path{doi:https://doi.org/10.1002/prot.26299}}.
\newline\urlprefix\url{https://onlinelibrary.wiley.com/doi/abs/10.1002/prot.26299}

\bibitem{RetroStat1}
S.~Watanabe, \href{https://link.aps.org/doi/10.1103/RevModPhys.27.26}{Symmetry
  of physical laws part i. symmetry in space-time and balance theorems}, Rev.
  Mod. Phys. 27 (1955) 26--39.
\newblock \href {https://doi.org/10.1103/RevModPhys.27.26}
  {\path{doi:10.1103/RevModPhys.27.26}}.
\newline\urlprefix\url{https://link.aps.org/doi/10.1103/RevModPhys.27.26}

\bibitem{RetroStat2}
C.~J. Ellison, J.~R. Mahoney, J.~P. Crutchfield,
  \href{https://doi.org/10.1007/s10955-009-9808-z}{Prediction, retrodiction,
  and the amount of information stored in the present}, Journal of Statistical
  Physics 136 (September 2009).
\newblock \href {https://doi.org/10.1007/s10955-009-9808-z}
  {\path{doi:10.1007/s10955-009-9808-z}}.
\newline\urlprefix\url{https://doi.org/10.1007/s10955-009-9808-z}

\bibitem{RetroStat3}
C.~C. Aw, F.~Buscemi, V.~Scarani,
  \href{https://doi.org/10.1116/5.0060893}{Fluctuation theorems with
  retrodiction rather than reverse processes}, {AVS} Quantum Science 3~(4)
  (2021) 045601.
\newblock \href {https://doi.org/10.1116/5.0060893}
  {\path{doi:10.1116/5.0060893}}.
\newline\urlprefix\url{https://doi.org/10.1116/5.0060893}

\bibitem{RetroQM1}
Y.~Aharonov, P.~G. Bergmann, J.~L. Lebowitz,
  \href{https://link.aps.org/doi/10.1103/PhysRev.134.B1410}{Time symmetry in
  the quantum process of measurement}, Phys. Rev. 134 (1964) B1410--B1416.
\newblock \href {https://doi.org/10.1103/PhysRev.134.B1410}
  {\path{doi:10.1103/PhysRev.134.B1410}}.
\newline\urlprefix\url{https://link.aps.org/doi/10.1103/PhysRev.134.B1410}

\bibitem{RetroQM2}
D.~Tan, S.~J. Weber, I.~Siddiqi, K.~M\o{}lmer, K.~W. Murch,
  \href{https://link.aps.org/doi/10.1103/PhysRevLett.114.090403}{Prediction and
  retrodiction for a continuously monitored superconducting qubit}, Phys. Rev.
  Lett. 114 (2015) 090403.
\newblock \href {https://doi.org/10.1103/PhysRevLett.114.090403}
  {\path{doi:10.1103/PhysRevLett.114.090403}}.
\newline\urlprefix\url{https://link.aps.org/doi/10.1103/PhysRevLett.114.090403}

\bibitem{RetroQM3}
S.~M. Barnett, J.~Jeffers, D.~T. Pegg,
  \href{https://doi.org/10.3390%2Fsym13040586}{Quantum retrodiction:
  Foundations and controversies}, Symmetry 13~(4) (2021) 586.
\newblock \href {https://doi.org/10.3390/sym13040586}
  {\path{doi:10.3390/sym13040586}}.
\newline\urlprefix\url{https://doi.org/10.3390%2Fsym13040586}

\bibitem{Kleinert2004-zx}
H.~Kleinert, Path integrals in quantum mechanics, statistics, polymer physics,
  and financial markets, 3rd Edition, World Scientific Publishing, Singapore,
  Singapore, 2004.

\bibitem{Del_Moral2004-jf}
P.~Del~Moral, Feynman-kac formulae, Probability and Its Applications, Springer,
  New York, NY, 2004.

\bibitem{Feynman}
R.~P. Feynman,
  \href{https://link.aps.org/doi/10.1103/RevModPhys.20.367}{Space-time approach
  to non-relativistic quantum mechanics}, Rev. Mod. Phys. 20 (1948) 367--387.
\newblock \href {https://doi.org/10.1103/RevModPhys.20.367}
  {\path{doi:10.1103/RevModPhys.20.367}}.
\newline\urlprefix\url{https://link.aps.org/doi/10.1103/RevModPhys.20.367}

\bibitem{Sivia2006}
D.~Sivia, J.~Skilling, Data analysis: a Bayesian tutorial, OUP Oxford, 2006.

\end{thebibliography}


%\begin{thebibliography}{00}


%\bibitem{}

%\end{thebibliography}
\end{document}